# Block Chain in the IoT industry: A Systematic Literature Review

**Kashif Ishaq[1], Fatima Khan[1]**

[1]School of Systems and Technology, University of the Management and Technology, Lahore, Pakistan

**Corresponding Author:** kashif.ishaq@umt.edu.pk

**ABSTRACT** The possibility of block chain innovation revolutionizing business operations and interpersonal interactions in Industry 4.0 is becoming more widely acknowledged. Industry 4.0 and the Industrial Internet of Things (IoT) are among the new application fields. As a result, the purpose of this article is to investigate the block chain applications that are already being used in IoT and Industry 4.0. In particular, it looks at current research trends in various IoT applications, addressing problems, concerns, and potential future uses of integrating block chain technology. This article also includes a thorough discussion of the key elements of block chain databases, including Merkle trees, transaction management, sharding, long-term memory, and short-term memory. In order to do this, more than 46 pertinent primary research that have been published in reputable journals have been chosen for additional examination. The workflow of a block chain network utilizing IoT is also demonstrated, demonstrating how IoT devices communicate with one another and how they contribute to the network's overall operation. The taxonomy diagram below serves to illustrate the contribution.

## I. Introduction:

In an industrial revolution, industry 4.0 brings up the latest phase that targeted interconnectivity, machine learning, automation, and real-time data. Sometimes this industry 4.0 is also known as IoT. The role of IoT and IoT is to interface various gadgets and bits of gear through the Internet, IoT and IoT can assist organizations with working all the more effectively, settle on more educated choices and open new income sources. A gathering of "things" installed by programming, gadgets, actuators, sensors, which are connected through the web to assemble and trade data with each other, the IoT nodes comprise sensor gadgets and handling energy to be available and common in different enterprises. Due to its potential influence on our daily lives and role in the development of smarter societies, the Internet of Things (IoT) is a dynamic and inventive technology that is growing in significance. Kevin Ashton first used the phrase "Internet of Things" (IoT) in 1999 to refer to linked sensors' capacity to introduce new services to the web [1]. Block chain (BC), on the other hand, is a distributed transaction and data management technology. Since 2008, there has been a steady increase in interest in BC technology. The BC development has had an influence, particularly in terms of tracking, organizing, carrying out, and storing data from several devices and offers data transparency, security, lineage, and provenance. Using

BC technology, transactions may be completed and carried out securely and with full traceability without the participation of a third party.

Smaller and smarter devices are being used in a variety of Internet of Things (IoT) applications, including supply chain management, smart home appliances, infrastructure management, agriculture, and healthcare. IoT devices have the ability to collect data in a private and secret manner, and there are several security-related dangers and flaws that aim to realize the IoT architecture of today [2]. Due to the price of hardware infrastructure and maintenance, IoT solutions are now in use. The IoT's scale and extensive use provide significant protection and security challenges. High-end servers with centralized network designs supporting the IoT are required by existing integration methods for providing protection, security, and information processing. In other words, when a centralized server fails, the entire network architecture becomes more vulnerable [3]. For instance, a single point of failure may emerge from a successful denial of service (DOS) assault on a centralized server.

Therefore, the creation of applications for the IoT-based platform might be problematic when using a centralized method. A major redesign of the IoT's architecture is necessary to address these issues. The Internet of Things can be supported by block chain technology, making it one of the finest solutions since it provides a secure, decentralized ledger maintenance mechanism [4]. Innovation from British Columbia has largely been adopted as the industry standard in a variety of industries, including banking and agriculture. In order to deliver "untrustworthy" functionality, the BC heavily relies on encryption. Data transaction nodes can coordinate more quickly since there is no centralized authorization network. The special abilities of the BC are used to create server less infrastructure.

## II.  Related Work:

IoT innovation has been generally embraced by the assembling business in machine-to-machine (M2M) correspondence. Here are some IOT applications [5] in fig1.

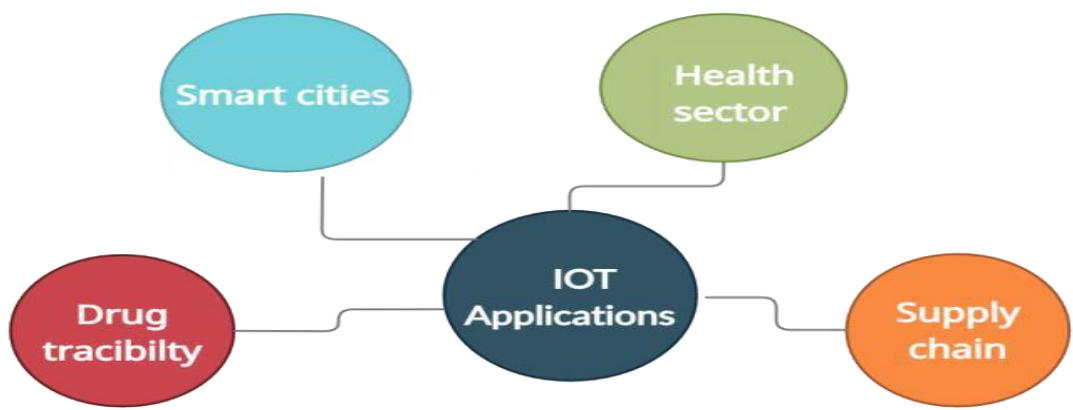

FIGURE 1.IoT applications.

Though the recent innovations make the idea of IoT possible, an enormous number of difficulties lie ahead for supporting the huge scope of genuine organization of IoT applications. Lately, the block chain has drawn in broad consideration from analysts and organizations for its security and straightforwardness. The block chain can be the authoritative construction for interlocking the whole lot and for time stamping varied information in Industry 4.0. The topic in this paper is the cooperation of BCT and the control of IoT [6]. Supposedly, research chips away at this topic are restricted because block chain is very notable in monetary administrations. In this part, we investigate those block chain advancements engaged with IoT by outlining a few ongoing examinations. A sum of 18 instances of block chain use have been sorted, and four of them are determined to IoT, including a permanent log of occasions and access the executives of information, detecting information exchanging, IoT gear exchanging, and IoT gadgets verification. The authors talk about the mix of the block chain with IoT and feature the combination advantages, difficulties, and future bearings. Basically, because of its decentralized elements in calculation and the executives' processes, the block chain can be a strong innovation to settle numerous IoT issues, particularly security. They elaborate that it is the correct method for moving the current concentrated IoT framework towards the decentralized design. To accomplish this point, the creators use an IoT door as a block chain hub and propose an occasion-based informing instrument for low-power IoT end gadgets. The authors survey the use of brilliant agreements in IoT and depict how shrewd agreements can work with and support the independent dividing of administrations between IoT gadgets. The importance that IoT can benefit from block chain networks is portrayed as far as exchanging, charging, shipment, and production network the executives. The creators present a discernibility framework for following Chinese agrifood supplies. The proposed framework consolidates the radio recurrence distinguishing proof with block chain innovation to improve food quality and wellbeing however in the meantime to decrease transportation costs. An IoT gadget the board framework is proposed to control and design IoT gadgets from a distance [7].

The creators propose an invigorating key administration conspire by which public keys are saved in Ethereum while private keys are put away on each IoT gadget. The Ethereum network is utilized in the verification of ideas since it gives the necessary resources to display brilliant agreement that can be run on top of the organization. Thusly, the upkeep and troubleshooting are improved since the update of code can straightforwardly change the conduct of IoT gadgets. A savvy city application system is proposed to incorporate heterogeneous shrewd gadgets in an exceptionally protected way.

The proposed structure gives an assortment of highlights, including better adaptation to internal failure capacity, further developed unwavering quality, versatility, and effective activity, which set up a typical block chain eco framework in which everything gadgets could speak with one another in a solid circulated climate. The creators propose lightweight engineering for IoT to wipe out the overheads of exemplary block chains while keeping up with most of its security and protection benefits. A private changeless record that is overseen midway is intended to advance energy utilization from IoT gadgets. Furthermore, they utilize the circulated trust to lessen the square approval handling time. Finally, a delegate contextual investigation indicated for brilliant home is executed to investigate the ease of use of the proposed design. Fair Access is a completely decentralized administration structure because of block chain that empowers clients to possess and control gadget information. In this structure, new sorts of exchanges are intended to issue and deny the entrance utilizing shrewd agreements. A decentralized, cloud-based stage determined for industry fabricating based on block chain innovation. The authors assemble a confided in delegate for exchanges among the clients to give on-request admittance to assembling assets. Single board PCs, like Raspberry, are used for correspondence between machines, the cloud, and the block chain network. The authors proposed a decentralized framework, permitting sensors to trade Bitcoins with information.

In order for the sensor to respond to the client by gathering information, the client must transmit an exchange (in the form of Bitcoins) to the sensor's location. The designers envision a layer hoarding system based on a block chain that would enable end users complete control over their devices in order to ensure device ownership. The authors present a decentralized framework to protect clients' security on IoT gadgets (Bluetooth low energy modules) by utilizing the Ethereum stage [7].

## III. Methodology:

Expanded security and dependability of submitted information inside the sensor organization could be accomplished on an application level. Along these lines, a lightweight, significant level correspondence convention given block chain standards was planned [8]. This paper uses the algorithm which is known as consensus comprises of initially setting up a cycle to approve, check, and affirm transactions, then, at that point, recording the exchanges in a huge conveyed catalog, making a square record (a chain of squares), lastly executing an agreement convention.

Subsequently, approval, confirmation, agreement, and changeless recording lead to trust and security of the block chain.

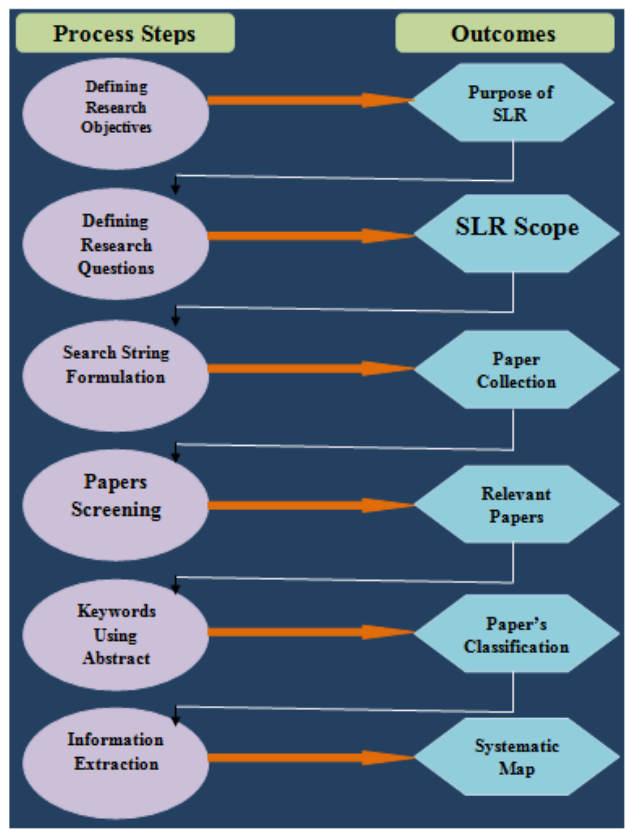

**FIGURE 2.** An SLR process model

The SLR outlines a step-by-step procedure for compiling, analyzing, and selecting the primary papers from all available research on the subject being evaluated. The requirements for SLR specified in this study were adhered to for objective data collection and demonstration of analyzed and extracted results.

Figure 1 shows this SLR's research methodology. This appropriate and mindful evaluation process has six levels: 1) Specific research goals must be established. 2) Specific research queries are outlined. 3) Selecting a search strategy 4) Paper selection and screening 5) Documents are categorized using keywords. 6) Data extraction and synthesis

## A. RESEARCH OBJECTIVES (RO)

Following are the main purposes of this study:

RO1: To investigate the optimal management and security of large-scale data using block chain technology.

RO2: To explore the use of block chain technology on Internet of Things (IoT) devices to improve data security, integrity, and administration in a decentralized and transparent manner.

RO3: To evaluate the benefits of block chain adoption and to pinpoint the implementation's difficulties.

## B. RESEARCH QUESTIONS (RQ)

The principal research topics have been defined in order to adequately perform this SLR. A detailed search strategy was created as part of the study in order to find and extract the most important publications. Table I includes a list of the research topics covered in this study along with their key objectives. The methods described in [24] and [25] are followed for addressing and responding to the inquiries.

**TABLE I.** RQ and major motivations

| Research Questions | Major Motivations |
|---|---|
| **RQ1:** Which are the publication channels on IoT and block chain also along with quality assessment? | Investigate IoT and block chain publication channels to stay current and ensure accurate information distribution through quality evaluation. |
| **RQ2:** How a large data will be managed and secure using block chain technology? | To effectively manage and safeguard vast amounts of data, decentralized, transparent, and impenetrable data storage and protection must be implemented. |
| **RQ3:** How block chain will implement on IoT? | To offer a decentralized, transparent, and secure framework for data management and exchange amongst IoT devices. |
| **RQ4:** Why block chain should be used and what are the challenges of using it? | To improve security, transparency, and trust in different applications, however difficulties include scalability, interoperability, and regulatory issues. |

The answer to these questions is **RA1:** By using a clustering method [9] and the canal (Private subnetwork), Block chain can be distributed into clusters with a block chain manager (BCM) per cluster sustaining one replica of the ledger and letting dissemination of someone's data personal and secure. The canal allows cooperation with each other within a network to achieve private transactions by maintaining a ledger available to members of the subnetwork. PBFT consensus mechanism is used to achieve energy efficiency in one's system. For the answer to **RQ2** see the below description and for **RQ3** see section I.

### *C. Implementation of block chain*

There are 3 domains in which block chain can be implemented [10].

    A.  Public: This type of network need no permission one can connect with the network and read or send the transaction. Anyone can be minor and do the consensus process without any consent.

    B.  Private: In this type of network No one can join the network without consent. This type of system is installed in an industry where no
one can send the transactions out of the network.

    C.  Consortium area: It comes under the limited authorization, just authorized hubs can be in the consensus process. The authorization to peruse or send might be unveiled or might be given distinctly to not many authorized hubs.

### *D. Implementation of IoT*

At first, this section describes the communication model (P2P messaging, distributed data sharing, and Autonomous coordination with the device) In any case, sadly, it has its impediments, which make it dangerous to implement:

**1:** Slow in Processing: For the most part, low-end CPUs are accessible which makes it hard for handling, as calculations in block chain require high CPU and memory to work appropriately.

**2:** Less Storage: As an ever-increasing number of exchanges are made the ledger's size continues expanding.

In any case, organizations like IoT A, like block chain small sensors, are proposing new methodologies: resulting in a lessening in equipment require necessities by working on the course of mining. Block chain network with the execution of innovation: mining, record, encryption, and so forth. Here, block chain hubs are the individuals from the network, participating effectively in the exchange interaction. They approve the exchange through mining; they can be PCs, endeavor servers, or likewise cloud-based hubs. Clients are IoT gadgets; they don't store the appropriate record. Block chain clients' square chain hubs communicate with one another through APIs. IoT gadgets make exchanges and these exchanges are handed off to block chain hubs for handling and putting away the information into the appropriate record. The HTTP REST APIs can be utilized to layout correspondence among IoT and block chain. They are explicit for each block chain hub. With the new investigation, it has been anticipated that the future will be developed by numerous block chains, with each having various highlights and giving unique administrations. Here, a Block chain organization might be a home organization, endeavor, or the web. Message designs and correspondence conventions between gadgets are out of the extent of block chain execution: it alludes to machine-to-machine correspondence. If man-made consciousness is added to the IoT climate that is associated with a block chain network it makes a Decentralize Autonomous Association (DAO), DAO alludes to an association that runs with practically no human mediation [11].

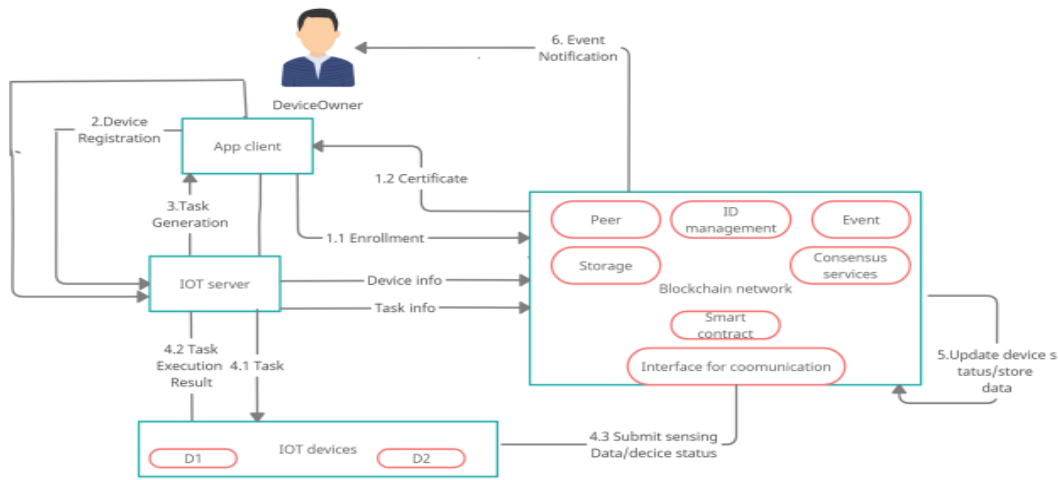

**FIGURE 3:** Structure workflow of IoT BCT platform.

### E. Configuration of IoT

Block chain innovation helps a great deal in laying out a trusted what's a more protected arrangement for IoT gadgets. Approaches that appear to be pertinent here are: Properties of IoT like Configuration subtleties and the last rendition firmware approved can be facilitated on the record. During bootstrap, the block chain hub is approached to get its arrangement from the record. The arrangement is expected to be encrypted in the record to forestall the revelation of IoT network geography or its properties by examination of the data put away in the public record.

| Components | Description |
|---|---|
| CPU | Intel core i5-8500@3.00 GHZ |
| Memory | 12 GB |
| OS | Ubuntu Linux 18.4.0 LTS |
| Docker Engine | Version 18.6.1-ce or latest |
| Docker compose | Version 1.13.0 or the latest |
| Node | V8.11.4 |
| Hyper-ledger Fabric | V1.2 |
| IDE | Composer-playground |
| SCLI tool | Composer-CLI, composer rest service |
| DBMS | Couch DB |
| Programming Language | Node.js |

**Table.2** Development environment **[13]**

The hash value worth of the most recent setup document for each gadget can be facilitated in the record. Utilizing a cloud administration the IoT gadget should down-stack the most recent and believed arrangement record after each decent time frame time (say consistently). Then, at that point, the gadget can utilize the block chain hub API to recover and match the hash esteem, which is put away in the block chain. This would permit the executives to eliminate any awful setups consistently and reboot every single IoT gadget in the organization with the most recent and trusted designs. Table.2 describes the developed IoT block chain network environment using hyper-ledger [12].

## D. SEARCH SCHEME

The creation of a search strategy is essential to successfully finding and gathering potentially important articles in the topic of interest. The description of a search string, the literary resources utilised to conduct the search, and the segregation (inclusion/exclusion) plan used to extract the most pertinent articles from the collection are all parts of this procedure. The gathered papers' many facets were evaluated qualitatively and experimentally to reflect a range of viewpoints related to the topic.

### Inclusion and Exclusion criteria

The inclusion and exclusion criteria for research in the context of IoT and block chain can be formulated as follows:

Inclusion criteria (IC):

IC 1) Consider research that primarily focuses on the application of IOT and block chain technologies.

IC 2) Include studies that aim to enhance or utilize IoT and block chain in the designated field.

To ensure relevance and specificity, the following exclusion criteria (EC) were applied:

Exclusion criteria (EC):

EC 1) Exclude studies that lack the integration of computerized components in their implementation of IOT and block chain.

EC 2) Exclude studies that do not provide a clear and defined methodology for incorporating IOT and block chain.

EC 3) Exclude studies that specifically concentrate on tracing the structural relationships of unrelated aspects.

EC 4) Exclude studies that primarily concentrate on binding mechanisms unrelated to the designated field.

By applying these inclusion and exclusion criteria, the research selection process will prioritize studies that primarily focus on the application of IoT and block chain technologies. Studies lacking computerized components, a clear methodology, relevance to the designated field, or focusing on unrelated aspects will be excluded.

### 1. Search String

By creating a keyword-based search string and utilizing it to find and compile the field's accessible studies across a number of well-known digital research repositories, an efficient and impartial study has been carried out. The main ideas have been examined in the context of the research questions in order to derive pertinent keywords and terms utilized in the chosen field of study. This has been done to ensure the validity of the search string regarding the relevance of its findings. The formalized search string needed to identify the most pertinent articles is described in Table III, together with the finalized keywords and their alternate phrases (synonyms). The '+' sign is used in Table III to indicate inclusion and the '-' sign to indicate exclusion of research using similar terminology.

**TABLE III. Terms and keywords used in search**

| Terms (Keywords) | Synonyms/Alternative keywords |
| --- | --- |
| + Block Chain (BC) | Distributed Ledger Technology (DLT), Digital Ledger (DL) |
| + Internet of Things (IoT) | Web of Things (WoT),  Smart System (SS) |

The finalized keywords and alternative phrases were combined to create a search string using the logical "AND" and "OR" operators.  When necessary, the wildcard "*" was also used to denote zero or more characters. While the "AND" operator joins the phrases to concatenate the terms to describe the search possibilities and to constrain the query to receive relevant search results, the "OR" operator provides for more search alternatives.

$$R=\forall\ [(CB \lor CL \lor ML) \land (PP \lor TP \lor PoP) \not\equiv (MS \lor SY \lor SC \lor SM \lor B)] \tag{1}$$

Here in (1), R stands for the search results obtained against the search phrase, with " denoting 'for all,' " being used for the 'OR' operator, and " being used for the 'AND' operator in combination with the search keywords specified in Table III to formalize the whole search string in accordance with each chosen repository. When utilizing (1), the general search word may be written as:

(block chain OR digital ledger OR distributed ledger technology) AND ("internet of things" OR "web of things" OR "smart system")

### 2. LITERATURE RESOURCES

Field specific and most prominent journals have been selected to conduct the literature search from online repositories, dedicated for research publication and collection. The details of selected repositories, applied search strings and the results are mentioned in Table IV.

**TABLE IV. Publisher wise search strings**

| Repository | Search String |
| --- | --- |
| PLOS | (("BLOCK CHAIN" OR DIGITAL LEDGER OR DISTRIBUTED LEDGER TECHNOLOGY) AND (INTERNET OF THINGS OR |
| | *WEB OF THINGS OR SMART SYSTEM))* |
| *ACM* | *("BLOCK CHAIN"[ALL FIELDS] OR DIGITAL LEDGER[ALL FIELDS] OR DISTRIBUTED LEDGER TECHNOLOGYALL FIELDS]) AND ("INTERNET OF THINGS"[ALL FIELDS] OR " WEB OF THINGS"[ALL FIELDS] OR " SMART SYSTEM"[ALL FIELDS] OR " INTERNET OF THINGS"[ALL FIELDS])* |
| **OXFORD ACADEMICS** | *("BLOCK CHAIN" OR DIGITAL LEDGER OR DISTRIBUTED LEDGER TECHNOLOGY) AND (INTERNET OF THINGS OR WEB OF THINGS OR SMART SYSTEM)* |
| *SPRINGER LINK* | *("BLOCK CHAIN" OR DIGITAL LEDGER OR DISTRIBUTED LEDGER TECHNOLOGY) AND (INTERNET OF THINGS OR WEB OF THINGS OR SMART SYSTEM)* |
| *SCIENCE DIRECT* | *(BLOCK CHAIN OR DIGITAL LEDGER OR DISTRIBUTED LEDGER TECHNOLOGY) AND (INTERNET OF THINGS OR WEB OF THINGS OR SMART SYSTEM)* |
| *IEEE XPLORE* | *((((("ALL METADATA":" BLOCK CHAIN ") OR "ALL METADATA": DIGITAL LEDGER ) OR "ALL METADATA": DISTRIBUTED LEDGER TECHNOLOGY) AND "ALL METADATA": INTERNET OF THINGS) OR "ALL METADATA": WEB OF THINGS)* |
| *PUBMED* | *("BLOCK CHAIN"[ALL FIELDS] OR DIGITAL LEDGER [ALL FIELDS] OR DISTRIBUTED LEDGER TECHNOLOGY[ALL FIELDS]) AND ("INTERNET OF THINGS"[ALL FIELDS] OR " WEB OF THINGS"[ALL FIELDS] OR " SMART SYSTEM"[ALL FIELDS] OR " INTERNET OF THINGS"[ALL FIELDS])* |
| *ACM DIGITAL LIBRARY* | *((((("BLOCK CHAIN" OR " BLOCK CHAIN") AND BASED) OR DIGITAL LEDGEROR (("BLOCK " AND " CHAIN") OR " BLOCK CHAIN")) AND ((("INTERNET OF THINGS" OR " INTERNET OF THINGS") AND ("WEB OF THINGS" OR " WEB OF THINGS") AND SMART SYSTEM) OR (("INTERNET OF THINGS" OR " WEB OF THINGS") AND SMART SYSTEM) OR ("INTERNET OF THINGSOF" AND ("BLOCK CHAIN" OR " DIGITAL LEDGER"))))* |

## F.SELECTIONOF RELEVENT PAPERS

The first search strategy produced duplicates and a sizable number of research publications, some of which are not directly related to the defined study subjects. As a result, to find publications that are actually relevant, a reevaluation and careful inspection of the retrieved papers are required. The articles should be grouped according to their titles, and any duplicates should be deleted. By carefully examining a sizable number of publications that were made accessible for this study and eliminating any that were irrelevant to the subject, a complete assessment was carried out. Following established criteria, the chosen papers were subsequently included in the ensuing review step.

## G. ABSTRACT BASIS KEYWORD

The main components of the work, the most relevant keywords, and its contribution to the issue were first determined by conducting an analysis of the abstract. By combining the key terms acquired from several articles, this technique helped to create a thorough understanding of how the study provides value to the field. As a result, these search terms were used to find the pertinent publications for the review process.

## V. Data Analysis

## A. ASSESSMENT OF RESEARCH QUESTIONS

## 1) ASSESSMENT OF QUESTION 1. WHICH ARE THE PUBLICATION CHANNELS ON IOT AND BLOCK CHAIN ALSO ALONG WITH QUALITY ASSESSMENT?

The internal criteria are used to evaluate an article's internal quality shown in the fig 4, while the stability and dependability of the publication source are taken into account to analyze an article's exterior quality the "computer science conference rankings (core)" and the "journal citation reports (jcr)" have been used to score and assess the external quality. The individual scores for each category are added to create the overall score. The final score can range from a minimum of 0 to a maximum of 10, and it can be classified as high ranked if it is larger than 8, ordinary ranked if it is between 6 and 8, and low ranked if it is less than 6.

| REFRENCES | PUBLICATION TYPE | PUBLICATION SOURCE | YEAR | INTERNAL SCORING | EXTERNAL SCORING | TOTAL SCORING |
|---|---|---|---|---|---|---|
| [1] | JOURNAL | SURVEY | 2018 | 0.5 | 0.5 | 1 |
| [2] | JOURNAL | PROBLEM BASED | 2019 | 1 | 0.5 | 1.5 |
| [3] | JOURNAL | PROBLEM BASED | 2013 | 1 | 1 | 2 |
| [4] | JOURNAL | SLR | 2018 | 1 | 1 | 2 |
| [5] | JOURNAL | PROBLEM BASED | 2019 | 0.5 | 0.5 | 1 |
| [6] | CONFERENCE | PROBLEM BASED | 2017 | 0.5 | 1 | 1.5 |
| [7] | CONFERENCE | PROBLEM BASED | 2017 | 0.5 | 0.5 | 1 |
| [8] | JOURNAL | PROBLEM BASED | 2017 | 1 | 1 | 2 |
| [9] | JOURNAL | PROBLEM BASED | 2018 | 1 | 0.5 | 1.5 |
| [10] | JOURNAL | SLR | 2020 | 1 | 1.5 | 2.5 |
| [11] | JOURNAL | PROBLEM BASED | 2021 | 1.5 | 1 | 2.5 |
| [12] | CONFERENCE | PROBLEM BASED | 2020 | 1 | 0.5 | 1.5 |
| [13] | CONFERENCE | PROBLEM BASED | 2016 | 0.5 | 0.5 | 1 |
| [14] | CONFERENCE | PROBLEM | 2018 | 1.5 | 1 | 2.5 |

| | | BASED | | | | |
|---|---|---|---|---|---|---|
| [15] | CONFERENCE | PROBLEM BASED | 2021 | 1 | 0.5 | 1.5 |
| [16] | CONFERENCE | PROBLEM BASED | 2013 | 0.5 | 0.5 | 1 |
| [17] | JOURNAL | PROBLEM BASED | 2018 | 1 | 1 | 2 |
| [18] | JOURNAL | PROBLEM BASED | 2018 | 1.5 | 1.5 | 3 |
| [19] | CONFERENCE | PROBLEM BASED | 2016 | 0.5 | 0.5 | 1 |
| [20] | CONFERENCE | PROBLEM BASED | 2017 | 1 | 1.5 | 2.5 |
| [21] | CONFERENCE | PROBLEM BASED | 2017 | 1.5 | 0.5 | 2 |
| [22] | CONFERENCE | PROBLEM BASED | 2017 | 0.5 | 0.5 | 1 |
| [23] | JOURNAL | PROBLEM BASED | 2020 | 0.5 | 1 | 1.5 |
| [24] | JOURNAL | SLR | 2004 | 0.5 | 0.5 | 1 |
| [25] | JOURNAL | SLR | 2017 | 0.5 | 1 | 1.5 |
| [26] | JOURNAL | PROBLEM BASED | 2019 | 0.5 | 0.5 | 1 |
| [27] | CONFERENCE | PROBLEM BASED | 2017 | 1 | 1.5 | 2.5 |
| [28] | CONFERENCE | PROBLEM BASED | 2017 | 0.5 | 1 | 1.5 |
| [29] | JOURNAL | PROBLEM BASED | 2014 | 0.5 | 0.5 | 1 |
| [30] | CONFERENCE | PROBLEM BASED | 2018 | 1 | 0.5 | 1.5 |
| [31] | CONFERENCE | PROBLEM BASED | 2018 | 1 | 1.5 | 2.5 |
| [32] | JOURNAL | SURVEY | 2021 | 1.5 | 1.5 | 3 |
| [33] | CONFERENCE | PROBLEM BASED | 2017 | 1 | 1.5 | 2.5 |

| [34] | JOURNAL | PROBLEM BASED | 2018 | 0.5 | 0.5 | 1 |
|------|---------|---------------|------|-----|-----|---|
| [35] | JOURNAL | PROBLEM BASED | 2018 | 1 | 0.5 | 1.5 |
| [36] | CONFERENCE | PROBLEM BASED | 2021 | 0.5 | 0.5 | 1 |
| [37] | CONFERENCE | PROBLEM BASED | 2015 | 0.5 | 1 | 1.5 |
| [38] | CONFERENCE | PROBLEM BASED | 2021 | 1.5 | 0.5 | 2 |
| [39] | JOURNAL | SURVEY | 2020 | 1 | 1.5 | 2.5 |
| [40] | JOURNAL | PROBLEM BASED | 2008 | 0.5 | 0.5 | 1 |
| [41] | CONFERENCE | PROBLEM BASED | 2017 | 1 | 1 | 2 |
| [42] | JOURNAL | PROBLEM BASED | 1997 | 0.5 | 0.5 | 1 |
| [43] | CONFERENCE | SURVEY | 2021 | 0.5 | 0.5 | 1 |
| [44] | JOURNAL | SURVEY | 2017 | 1.5 | 1 | 2.5 |
| [45] | JOURNAL | PROBLEM BASED | 2018 | 0.5 | 0.5 | 1 |
| [46] | JOURNAL | SURVEY | 2014 | 0.5 | 1 | 1.5 |

*Table.3 Quality Assessment Table*

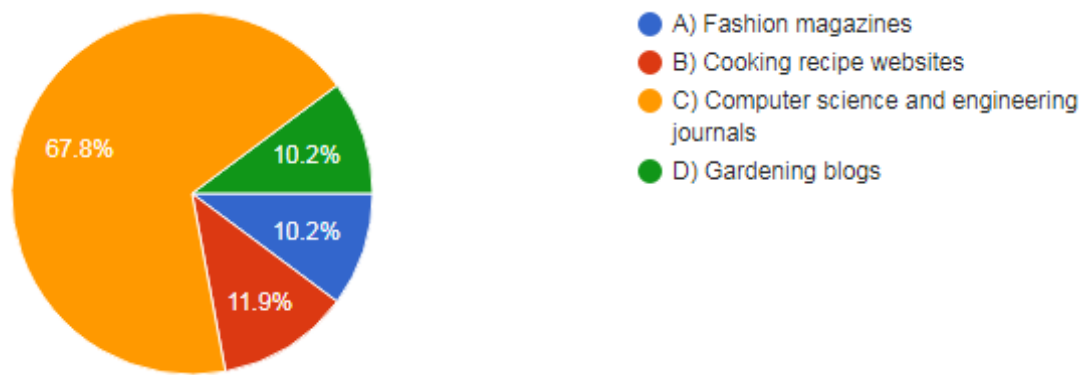

- A) Fashion magazines
- B) Cooking recipe websites
- C) Computer science and engineering journals
- D) Gardening blogs

**2) ASSESSMENT OF QUESTION 2. HOW A LARGE DATA WILL BE MANAGED AND SECURE USING BLOCK CHAIN TECHNOLOGY?**

## A) BLOCK CHAIN

Satoshi Nakamoto made block chain famous with Bitcoin in 2008 **[26]**. To avoid duplicate spending, block chain technology was initially used to build Bitcoin. But today, block chain is being used for many other things, as the IoT in this research shown in the fig 7 given below.

The word "block chain" is frequently used to refer to data structures, networks, or other technologies. Each block in a block chain is an ordered collection of transactions. A hash from the previous block is used to connect each block in the block chain to the one before it. As a result, the block chain's transaction history is unchangeable since doing so would entail modifying or erasing the whole block chain's data **[27]**. The immutability of the block chain is what protects it from hackers.

The block structure and geometry of the block chain are as follows.

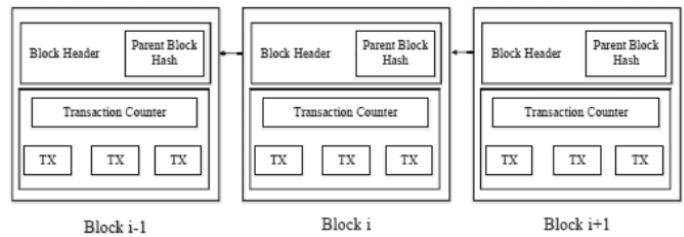

**FIG NO.5** Block Chain **[28]**

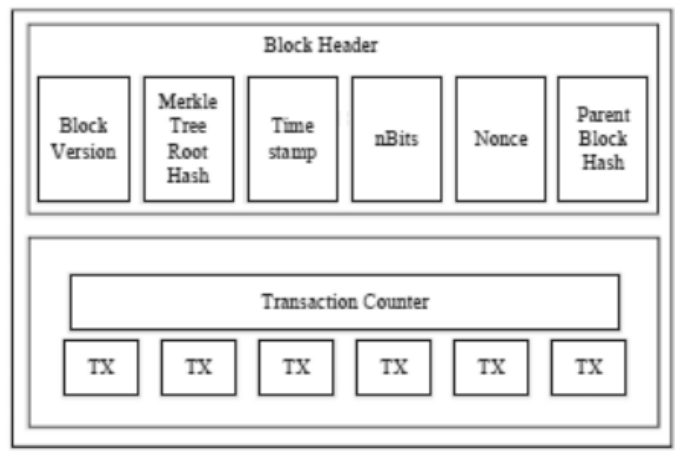

**FIG NO. 6** Block Structure **[28]**

Several crucial traits of block chain exist **[28]**:

• **Decentralization:** In a block chain, transactions may be verified by all parties involved. Instead, to guarantee data consistency throughout the network, consensus procedures are used.

• **Persistence:** Transactions on the block chain can be confirmed quickly, and miners will reject any transactions that are invalid. Consequently, once a transaction has taken place, it cannot be deleted.

• **Anonymity:** On the block chain network, any user may communicate with others using a randomly created address, protecting their true identities.

• **Auditability:** Every transaction on the block chain makes a reference to the one before it, making it simple to verify and follow the history of each transaction.

**B) ETHEREUM**

The Vitalik Buterin-developed block chain platform Ethereum overcomes some of Bitcoin's drawbacks [29]. One of Ethereum's key benefits is its support for full Turing-completeness, which enables all forms of computing [30]. It acts as a state machine with a transactional basis [31].

These fundamental components are included in ethereum [31]:

• **Currency:** The "Ether" or ETH intrinsic currency used by Ethereum is used for network computing and data transfer.

• **Account:** The four components of an Ethereum account are the nonce counter, storage, ether balance, and contract code. Each account has a 20-byte address. In Ethereum, there are two different kinds of accounts: Externally Owned Accounts (EOA), which are managed by private keys, and Contract Accounts, which are managed by contract code. Only EOA is capable of activating Contract Accounts.

• **Transaction:** A signed data package containing messages to be delivered from EOA is referred to as a transaction in the Ethereum protocol. Transactions come in two flavors: message calls and account creation. Each transaction comprises the sender's signature, the destination of the message, the quantity of Ether and data being transmitted, the number of Start Gas (Gas limit), and the price of Gas.

• **Technology Used:** Ethereum makes use of a number of technologies, including as data storage, client/node implementation, and web technology.

• **Consensus Algorithm:** Ethereum uses the Proof of Stake (PoS), Proof of Authority (PoA), and Proof of Work (PoW) consensus algorithms.

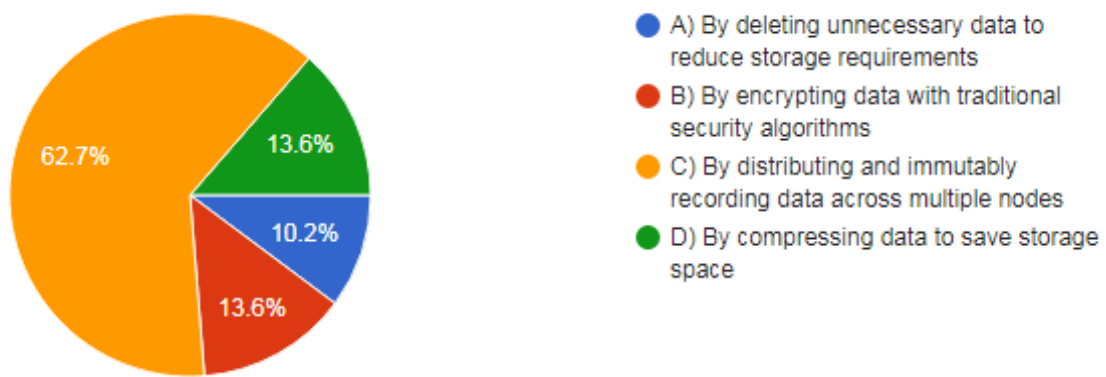

**FIG NO. 7 MANAGED AND SECURE USING BLOCK CHAIN TECHNOLOGY**

## 3) ASSESSMENT OF QUESTION 3. HOW BLOCK CHAIN WILL IMPLEMENT ON IOT?

IoT block chain implementation is a fascinating field of study that has attracted a lot of interest recently. In order to improve data security, privacy, and dependability, the integration of block chain technology with IoT devices has been studied in a number of research publications. The following is a synopsis of how block chain may be used on IoT devices, backed up by pertinent academic papers shown in fig 8:

**Decentralized Data Management:** The decentralized structure of block chain means that data produced by IoT devices is disseminated over a network of nodes rather than being held in a central repository. As a result, data redundancy is improved and the chance of a single point of failure is reduced [32].

**Data Integrity and Immutability:** The immutability and tamper-resistance of the block chain ensure that data from IoT devices cannot be changed or removed once it has been stored there [33]. As a result, data integrity is guaranteed, which is essential for IoT applications like supply chain management and healthcare.

**Secure Data sharing:** Block chain enables direct, peer-to-peer IoT device data exchange that is both secure and private, doing away with the need for middlemen and improving data privacy [34].

**Automation Using Smart Contracts:** Block chain-based smart contracts allow for the automatic execution of established rules and conditions, easing IoT device interactions and requiring less human involvement [35].

**Consensus Mechanisms:** The use of consensus mechanisms in block chain technology guarantees that information provided by Internet of Things (IoT) devices is reviewed and approved by the network before being added to the block chain, assuring data integrity and reliability [36].

**Identity and Access Management:** Block chain technology may be used for IoT identity and access management, offering a safe and decentralized method of device authentication [37].

**Scalability Solutions:** Several academic publications suggest sharding, side chains, and off-chain data storage as solutions to the scalability issues posed by block chain [38].

**Interoperability:** Research focuses on creating protocols and standards to enable seamless data sharing, and ensuring interoperability between various block chain networks and IoT devices is a significant aspect [39].

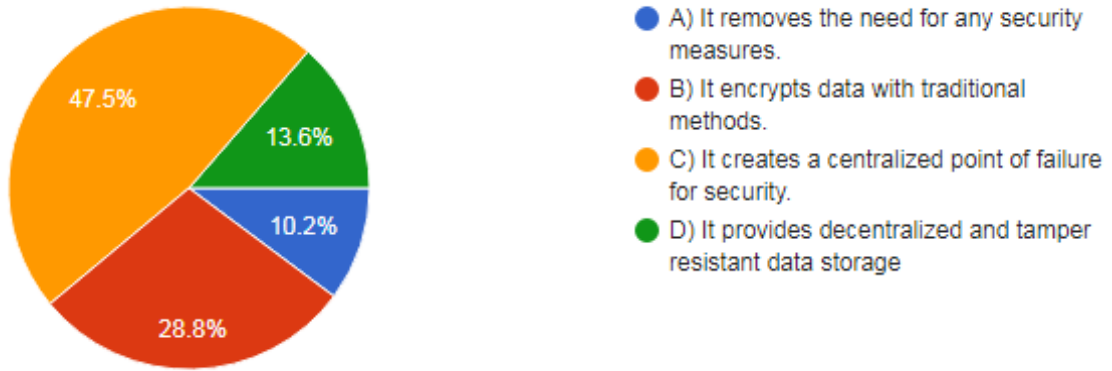

## 4) ASSESSMENT OF QUESTION 4. WHY BLOCK CHAIN SHOULD BE USED AND WHAT ARE THE CHALLENGES OF USING IT?

Block chain technology is an ideal choice for many applications since it has a number of strong advantages. the following factors support the use of block chain, according to studies shown in fig 9:

**Immutability and data integrity:** Data is immutable and tamper-resistant once it is stored on the block chain, assuring data integrity [40].

**Transparency and audit ability:** The transparency and public verifiability of block chain technology enable all parties to observe and audit the data, encouraging trust and accountability [41].

**Smart contracts:** Block chain makes it possible to utilize smart contracts, which are self-executing contracts with established rules. Smart contracts automate operations and lessen the need for middlemen [42].

Although block chain has many benefits, there are also substantial problems, as shown in academic papers:

**Scalability:** Block chain networks may experience scalability problems as transaction volumes rise, resulting in longer processing times for transactions and greater costs [43].

**Energy use:** Some consensus processes, such proof of work (pow), demand a lot of processing power, which results in significant energy use and environmental issues [44].

**Interoperability:** It is difficult to ensure compatibility and interoperability across various block chain networks and protocols, which obstructs frictionless data interchange [45].

**Regulatory and legal challenges:** The block chain's decentralized and pseudonymous nature poses regulatory and legal issues, notably with regard to data privacy and adherence to current regulations [46].

**Governance and consensus:** To maintain equitable and productive participant consensus in block chain networks, decision-making and governance systems must be carefully considered **[47]**.

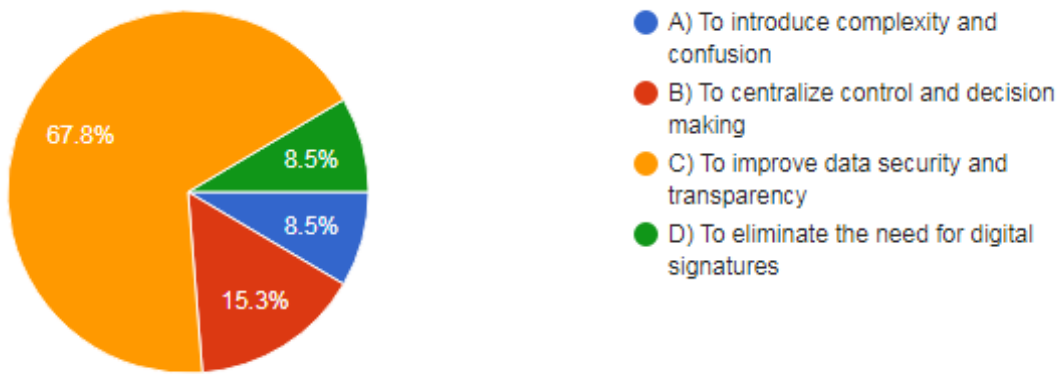

**FIG NO. 9 CHALLENGES OF USING IT**

## VI. Block chain technology overview

BCT is a chain of blocks that comprise data and information. In a block chain network, each block stocks some info as well as the hash of the prior block. A hash is a one-of-a-kind mathematical puzzle that detects a specific block. If the files within the block are reformed, the hash of the block will be reformed as well. Block chain is safe because it connects blocks using unique hash keys. While transactions take place on a block chain, the network's nodes validate them. These nodes are known as miners on the Bitcoin block chain, and they process and validate transactions on the network using the proof-of-work principle. Each block must refer to the hash of the block before it for a transaction to be legitimate. Only if the hash is correct will the transaction take place.

If a hacker tries to break into the network and change the information of a specific block, the hash associated with that block will be changed as well. Block chains are decentralized in nature, which means that the total network is not controlled by a single person or group. While everyone in the network has a copy of the distributed ledger, no one can change it independently. This one of-a-kind property of Block chain enables transparency and security while also providing people control.

The interaction between two parties through a peer-to-peer model is readily performed with the implementation of Block chain, and no third party is required. Block chain is based on the peer-to-peer (P2P) protocol, allowing all network participants to have identical copies of transactions, allowing for machine consensus approval. For example, if you want to perform a transaction from one part of the world to another, you may do so in a matter of seconds using Block chain. Furthermore, any delays or extra costs will not be subtracted from the transfer. The immutability attribute of a block chain relates to the idea that data written on the Block chain cannot be modified after it has been committed. Consider sending an email as an example of immutability. You can't take back an email you've sent to a group of people. You'll have to ask all your recipients to delete your email, which is a time consuming process to get around this.

As BCT has the feature of immutability, it is easier to identify data manipulation. Because every alteration in even one block can be recognized and handled smoothly, block chains are considered tamper-proof. Hashes and blocks are the 2 most common mechanisms to detect tampering. Each hash function linked with a chunk is unique, as earlier stated. You can think of it as a block's fingerprint. Any change in the data will cause the hash function to change. Because the hash function of 1 block is connected to the hash function of the succeeding block, a hacker would have to change the hashes of all the blocks after that block, which is challenging to perform. Block chain can be depicted as an information structure that holds conditional records. It can also be described as a data structure that has transactional records while ensuring security, transparency, and decentralization. Moreover, on a block chain, each transaction is protected by a digital signature that verifies its legitimacy—the data saved on the Block chain due to the encryption and digital signatures.

Block chain technology assists all network accomplices to create a consensus, also known as agreement. All files saved on a block chain are numerically recorded and have shared antiquity that is open to all network participants. This eliminates the possibility of any fraudulent conduct or transaction repetition without using a third party. **Transactions** in the Merkle tree (Describes in Section A) within a block are generated by digital fingerprint [18].

*Merkle Tree*

Merkle tree crossing requires just Log space and time1. On behalf of a tree through N hubs, our calculation processes successive tree and verification way information in time $Log2(N)$ & space under $3Log2(N)$, somewhere the units of calculation are hash assessments or leaf esteem calculations, and the components of the room are the number of hub values put away [15].

Comparative with this calculation, they demonstrate our limits to be fundamental and adequate. This outcome is an asymptotic improvement over any remaining past outcomes (for instance, estimating cost = space-time). They likewise demonstrate that the intricacy of our calculation is ideal: The layer structure [16] of block chain technology:

*A. Data Layer*

This level comprises transactions, hash functions, Merkle tree, blocks, and digital signature which refers to the approach of cryptographic algorithms. In block chain which is a chain of blocks, the first block is known as the genesis block which is linked to the new blocks in a chain. Each block in the chain has 2 parts transaction records and a header. Transaction records are structured in the Merkle tree. A Merkle tree alludes to a binary tree structure that sums up and permits content to be checked proficiently and safely inside an enormous information set. On the off chance that the exchanges are not pressed into Merkle trees, every hub of the organization would have to keep a total duplicate of every exchange which has at any point occurred on the block chain. Merkle trees are created by hashing hub matches over and over until only one hash is left and this hash is known as the root hash or the foundation of the Merkle tree. Each leaf hub holds a hash of exchange information, and each non-leaf hub contains a hash of its past hashes. An exchange is made when a client completes activities on the block chain. In all-purpose, the header of the block contains a hash of the preceding block for authentication, the root of a Merkle tree is used in transactions, a nonce for producing the hash value with the use of a consensus algorithm, and a timestamp alluding to the time when the block has been made.

*B. Consensus Layer*

No concentrated body is dispatched to screen the exchange or keep assailants from controlling or changing information when a hub trades information on the block chain network. To stay away from extortion-related exercises, for example, double-fold spending assaults, the reliability of the block should be checked, and the information stream ought to be controlled to guarantee the smooth trade of data. To avoid fraud block chain uses some protocols for validation purpose that refers to consensus algorithms. Proof of stake, proof of work, byzantine fault tolerance, proof of elapsed time, and proof of authority are some mechanisms that are used in consensus algorithm.

### C. Network Layer

P2P network is also known as a network layer that makes communication is possible between nodes. A P2P network is a mesh of nodes that ensures the data is distributed among all the computers in a network, and the responsibility of the network is shared across various hubs to accomplish the end target hubs on the block chain for handling blocks and transactions. Light node and full node are maintained by a P2P network. A light node can store only the deader of the block and a full node store the ledger. For the portion or distribution, block chain technology uses sharing techniques in a P2P network.

### Sharding:

In this strategy, a few block chains called a chain of a shard is overseen by network hubs as opposed to keeping a solitary block chain for all exchanges. Every shard comprises its hubs or validators that apply proof of work or cast a vote with a consensus mechanism.

### D. Infrastructure Layer

This paper describes the infrastructure layer of BC technology for 2 block chain enterprises: Hyperledger fabric [10] (HF) and Ethereum. A client's PC can take an interest in Ethereum block chain by running client programming like Geth, Parity, or Pantheon. Ethereum keeps up with two sorts of hubs: light hub, and full hub. The light hub runs the client programming stores the reserve and the condition of the Ethereum. Further, the light hub takes part in confirming the execution of exchanges while the full hubs download the whole record in their nearby stockpiling, take part in full agreement authorization,

Confirm signatures, exchanges, and block configurations and look at double-fold spending. The Ethereum hubs execute the Ethereum Virtual Machine (EVM) which resembles Java Virtual Machines (JVMs) running byte code. EVM going about as sandboxes offers an execution climate for a smart contract. EVM is Turing complete programming; a stake-based virtual machine that handles the interior state and calculation for smart contracts. The HF block chain is contained three sorts of hubs: **i)** endorsers, **ii)** orderers, and **iii)** peer hubs. The companion hubs have records and chain code that is otherwise called smart agreements. The utilizations' applications also managers utilizing Fabric Software Development Kit (SDK) APIs can constantly speak with peer hubs to get to the chain code or circulated record.

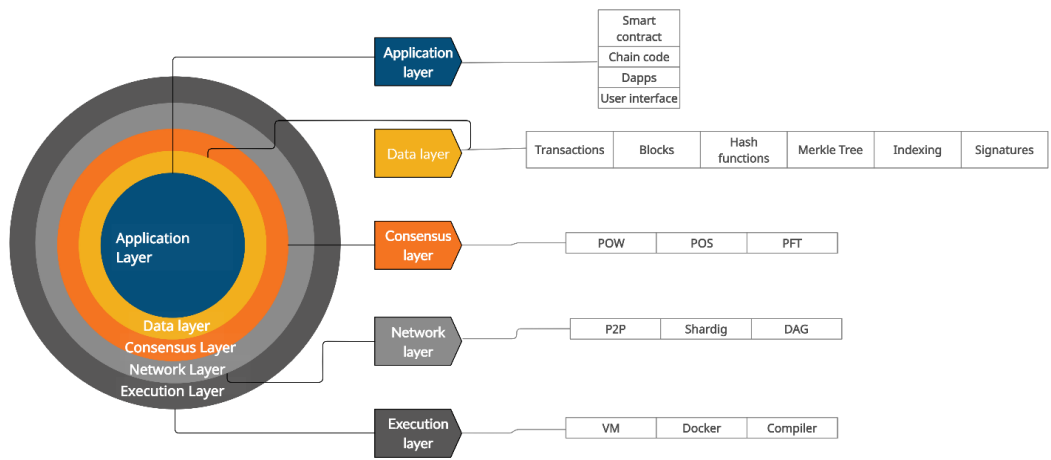

**FIGURE 10.** **Layers of Block chain technology.**

### E. Application Layer

This layer contains chain code, smart contracts, and what's more Apps. This layer involves 2 sub-layers show layer and the execution layer. The show layer incorporates contents, APIs, and client interface. These devices are utilized to associate the application layer with the block chain network. The execution layer incorporates brilliant agreements, chain code, and fundamental standards. The show layer sends directions to the execution layer, which runs exchanges. For instance, directions are shipped offchain code in HF and savvy contracts in EVM.

## VII. Block chain application in IoT:

Many kinds of applications such as healthcare, drug traceability, financial sector, supply chain, and many more in IoT can be permitted with block chain to maintain higher speed, security and minimize time. This paper describes these applications in detail.

### A. Supply chain

In human lives, SCM plays a vital role. Earlier, for supply management traditional system was used in which the order date, sending date, customer name, and address were stored there were some gaps in using the traditional database system they can't be able to trace data. Due to some security issues, the solution block chain is introduced in supply chain management. This technology in IoT makes life easier it can be able to provide transparency, less cost, and risk across the supply chain. Block chain technology in the supply chain can be beneficial in improving traceability of material supply chain to confirm business standards are met, lower losses from fake market trading, increase visibility and agreement over subcontracted contract trade and it also reduce the paperwork and organizational costs. But the question here is how the block chain technology assist supply chain management does? The answer to this question is in supply chain management systems it is hard to track the whole transaction process by using block chain when the transaction is made it is sent to all the people in the block chain they all have a copy of the ledger in distributed systems the ledger shows the time and transaction with the pubic id of a person. For the agreement or contacts between corporations, they use smart contracts for the signatures block chain has the feature of digital signatures which uses cryptographic algorithms (Hash algorithms). The minors in an organization or a network will verify all the operations so this way block chain can be implemented and make the data immutable, secure, visible, and distributed.

Let's study the proposed models for BCT. BPM business process management which is presented by Guerreiro et al. [16] This model uses an enterprise operating system (EOS) with block chain technology which reduces the risks in business transactions by expanding traceability, authenticity, and trust in contradiction of fraud. Through Leng etal [17] the agricultural supply chain was proposed this method was reliant on the dual chain structure that enhanced the effectiveness of the block chain in this system. The suggested result by the author is to deliver rent seeking plus corresponding tools for open service platforms. These explanations guaranteed visibility, privacy, and safety of confidential information. This system has some problems in contradiction of the performance and size of the BC. Mao [14] developed an architecture of credit evaluation which is also based on public BC in this architecture the text of the credit evaluation from brokers is collected by smart contracts and then a Long short term memory method is used to evaluate the text so this structure develops the effectiveness in management and supervision for food supply chain but this model is overpriced so all these problems were still in open block chain models related to budget, design and development. Subsequently, the best solution is implemented using block chain that is up to emphasizing in terms of both cost-effective and client fulfillment.

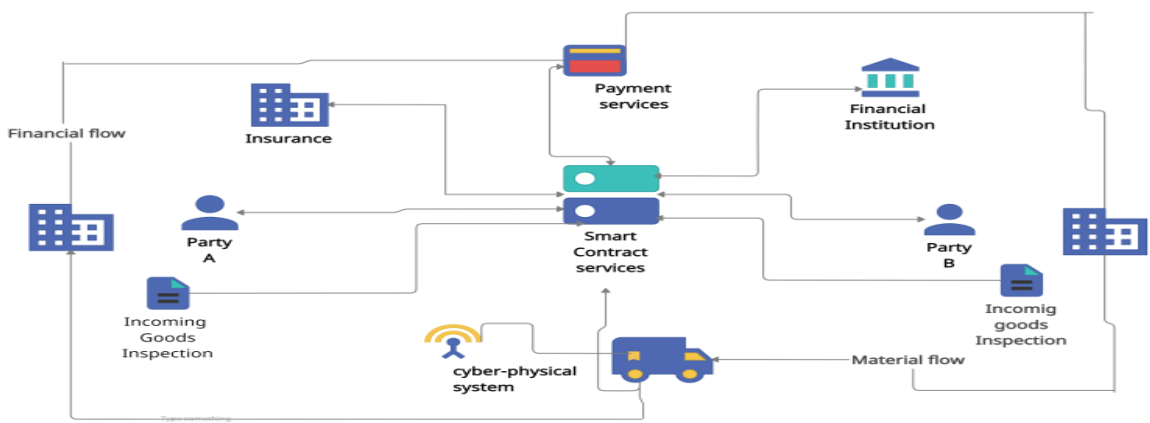

**FIGURE 11.** **Supply chain management using Block chain IOT.**

A double block chain system was presented by Leng et al. [17] it is mainly focused on storage management, supply rent-seeking, consensus algorithms, and matching mechanisms. Now the agricultural SC based on a twice chain structure can take care of the safety of transactions and privacy of the information of the enterprise. This model improved the reliability of the open platform facility and the general effectiveness.

## B. Healthcare sector

Health is the most important resource of any country. In conventional medical care, all understanding related information is put away in a unified way and hence, it isn't prudent to give access to information to any untrusted outsider. Additionally, the protection and security of patient data should be kept up with as it is helpless to an assortment of assaults. Concentrated engineering can't satisfy these prerequisites. Subsequently, savvy medical care is acquainted with managing the above notice issues in customary medical services frameworks. Excellent medical services center on checking and diagnosing patients' wellbeing somewhat through remote correspondence channels. It gathers the vital data of different patients over assorted wearable gadgets & sensors. Colossal information is gathered for enormous patients. It is a major concern to examine and collect this information in a got way. This information ought to likewise be shared safely among confided in gatherings like clinics, patients, specialists, and clinical stores. Secure correspondence of this information is significant as it influences significant choices such as arranging of new administrations in the clinic, suggesting specialists, breaking down manifestations of various sicknesses or medical problems, and working on the general framework to make it insightful.

Treatments, and examining indications of illnesses, patients' info is mandatory to be shared occasionally. Old-fashioned access mechanism rules are not safe enough to share very complex patient histories from one party to another. Furthermore, in most circumstances, patients do not share their health history with the doctors. In the situation of a medical crisis, medical accounts of the patient are needed, but the same is not open due to poor record maintenance. All the above-stated concerns can be unraveled by smart healthcare via Electronic Health Records (EHR). Although smart healthcare can resolve the main concerns in the healthcare industry, there are some challenges. Policy for EHR, confidentiality, safety, and accessibility, are open concerns in healthcare. BCT can be conditioned to get solutions to these concerns.

The block chain has the possible ways to upkeep smart healthcare through the dispersed ledger among numerous customers such as patients, clinicians, medical supplies, and insurance agencies. Healthcare data gateway HDG is a block chain-based intelligent application. With this secured HDG patients can steadily share and control the information it processes and achieves info of patients deprived of any concerns about the third party. MedRec [19] a decentralized model is proposed by Azaria et al [19].Which is an unusual data management system for huge scale HER. This structural design needed 4 new modules: BC-based distributed computing component towards examine parallel computing using data analytics and parallel, reliable info membership component to permit a reliable medical data system for cooperative research, application data management module to continue the integrity of data, and to integrate difference of health data, demonstrable unidentified identity management module for keeping identity secrecy for IoT, Person and devices also secure access to patient-centric medicine information. Rifi et al. [20] Discussed the vital glitches such as scalability, interoperability, and benefits of using block chain models in this sector to accomplish the best performance also discussed. In contradiction of these issues, the new model was recommended by Magyar et al [21] which is fundamentally integrated composite Electronic Healthcare Data EHD. This model uses a cryptographic approach for securing data. Jing et al [20] concentrated on the way that an ordinary immense number of medical services-related information is delivered by people what's more emergency clinics. This information is very helpful in the clinical industry for different purposes. It is very difficult what's more vital to store this information safely. It is a must to have systems to keep up with privacy, security, and uprightness of information. By remembering these prerequisites, they developed the Blochie model as block chain healthcare information exchange which has numerous ways and conditions to share and collect the information related to healthcare. The figure for the healthcare sector in IoT block chain is given below (Figure 4).

### C. Smart City

Recent smart areas are being changed through the mix of IoT advancements and stages into genuine smart cities. To acknowledge such a vision, brilliant city administrations [22] require reliable information from countless heterogeneous sources. Therefore, the environment of distributed block chain and the assurance of data variability and immutability could serve as a core component of more secure and trustworthy information-driven services, where protection perspectives should be appropriately addressed. In the Implementation of a smart city, assorted sensors [23] are used by several smart devices and consumers to gather the required information. This information is processed and used in the management of traffic, transportation, schools, and water resource networks, waste managing, public services, and power strategies to enhance performance.

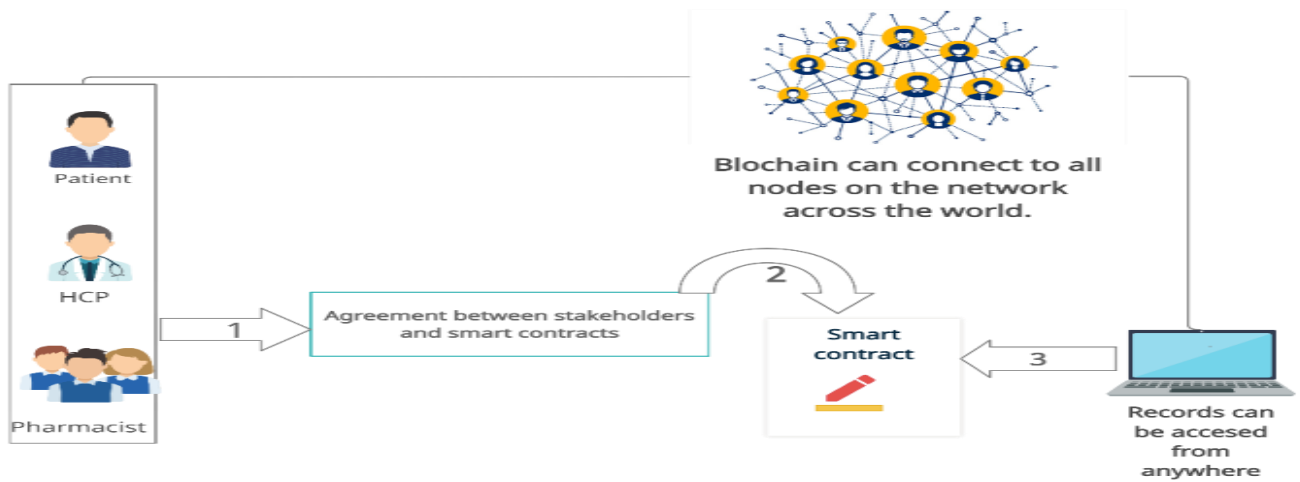

**FIGURE 12. Healthcare sector in IoT Block chain.**

## VIII. Section I:

**Reasons for using block chain:**
**Centralization:**
Traditional central architectures are more difficult for an organization if the central system gets attacked all the data will lose. The central system can be replaced with a decentralized BC for avoiding attacks and data losses.
**Trust establishment:**
These days establishing trust is the major problem with the third party for rising issues with money fraud or scam. The BC can be used to prevent these types of issues with the help of a ledger.
**Challenges of block chain:**
**Scalability:**
These days, people store every one of the exchanges in the decentralized block chain system to approve them. Accordingly, the block chain turns out to be weighty and slow. There is a limitation on the size of each square and the time expected to make another square. The block chain can handle simply 7 to 8 exchanges in one moment, however progressively situations, a large number of exchanges executed, and subsequently, it is difficult to carry out a block chain for a constant frame situation which forces a test of adaptability.

**Processing time and power:**

Expected to perform encryption calculations for every one of the items associated with Block chain-based IoT system given the way that IoT environments are extremely assorted and contained gadgets that have altogether different registering capacities, and not every one of them will be equipped for running similar encryption calculations at the ideal speed.

**The taxonomy diagram below serves as the contribution.**

Diagram of the taxonomy

The categorization of block chain applications in the IoT sector is shown in the diagram. Aspects including applications, block chain platforms, consensus processes, privacy and security, interoperability, scalability, governance, integration difficulties, and regulatory and legal issues are all categorized.

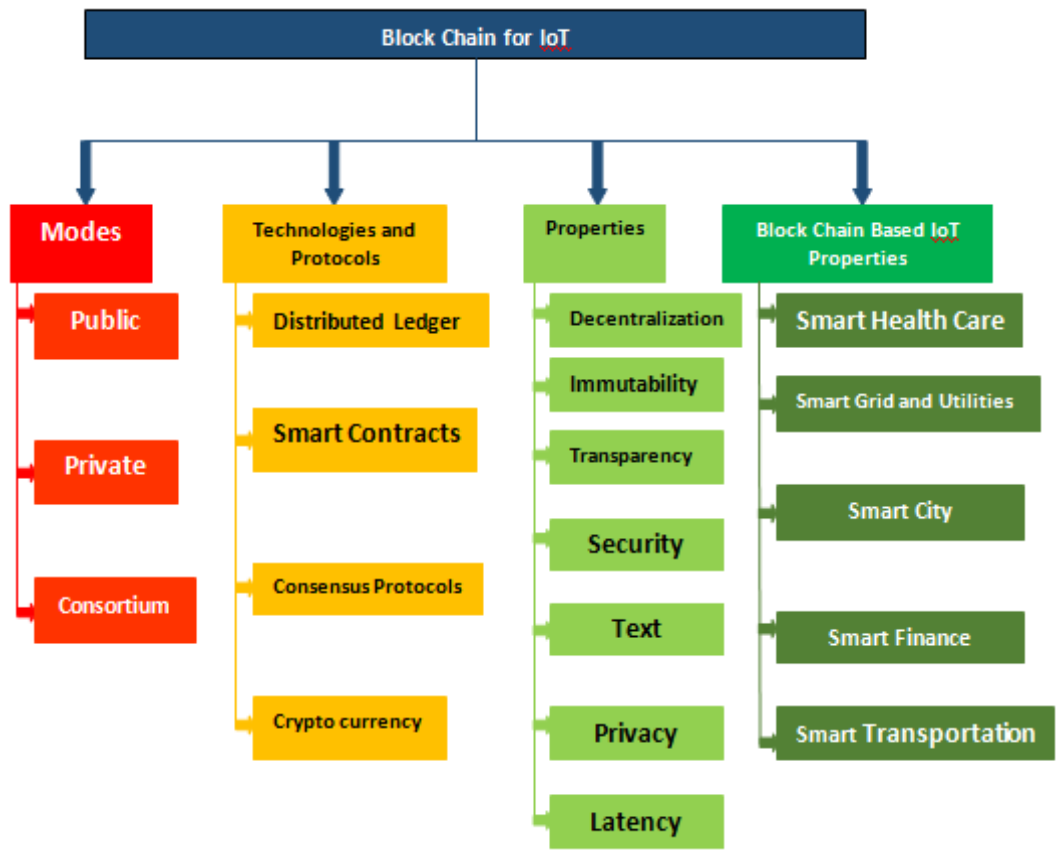

**Figure no. 13** Taxonomy or Block Chain in IoT

## IX. Conclusion

The utilization of block chain innovation in IoT empowered ventures (Industry 4.0 and Industrial IoT) is relied upon to develop and benefit a wide scope of modern areas. Carefully designed what's more strong procedures of record-keeping are supposed to be the distinct advantages presented by this innovation. As seen in various contextual investigations assessed in this paper, there have been huge interests in the block chain. Regardless of adaptability issues, the conventional strategy for concentrated information sharing has been viewed as more secure contrasted with a decentralized methodology, furthermore, has been practically speaking in the business for such a long time now. This is set to change now with block chain promising arrangements for tending to both security and respectability issues. In this paper, we reviewed the most recent exploration work directed on block chain materialness in numerous IoT explicit businesses. We likewise investigated different business executions of block chain in the industry 4.0 and IoT to give a dynamic proportion of reception by and by. Furthermore, we further examined the difficulties looked at by every one of these ventures for carrying out block chain. While numerous block chain projects have arisen over the most recent couple of years, research in the block chain is in its earliest stages. For block chain to be completely utilizable and adjustable, industry-arranged examination ought to be additionally coordinated to address large numbers of these difficulties counting individual information insurance, adaptability of square information, information mystery and security of the taking part association, block chain joining and reception cost, and administrative guidelines.

# X. Future Work

## A. Music industry

A brilliant agreement guarantees the legitimacy of a protected item. These agreements characterize and mechanize the connections and cooperation between the partners of an item. They additionally help in guaranteeing that the item a client is purchasing is real and in addition to an ideal duplicate. A potential use instance of brilliant agreements in the music business is the music privileges the executives. With the ascent of the Internet and mushrooming of a few web-based music real-time features, the music business has been going through a significant change in the last 10-15 years. Musicians, specialists, distributors, real-time feature suppliers, and numerous others engaged with this industry have been impacted by this change. The technique for eminence assurance in the music business has been all the time a confounded errand, however with the Internet rise, this cycle has become considerably more perplexing. Block chain innovation can help to make the sovereignty installment process straightforward by keeping an exact and complete decentralized data set for putting away data (as a public record) of music freedoms possession. It would get rid of the intermediation of the business, permitting craftsmen to make and catch additional worth from their items. Also, to not entirely set in stone by the shrewd agreements the eminence split for each work could be added to the record in this way ensuring a fair eminence offer to the first partners of the protected item.

## B. Educational industry

As the trend of online distance learning increments, so does the requirement for a free and straightforward method of confirming instructive records and records of understudies. A block chain-based stage could fill in as a public accountant for instructive records, making a way for instructive organizations and businesses to get records and records. It could likewise empower the coordinated effort of colleges and other scholastic organizations. The undeniably powerful nature of business infers that more capabilities are expected as vocation ways digress across disciplines and associations. Frequently capabilities are essential just to apply for a job, and with the ascent sought after, instructive untrustworthiness is likewise expanding block chain can turn into the administrations to the inspiration of dispensing with the broker in question and in saving the driving privileges. In contrast to the negative encounters of the current administrations because of a brought together model, the drivers can encounter total independence. The drivers in this assistance are allowed to talk with the clients, fix their charges, pick the installment strategy for their decision and work together with another driver.